\documentclass[12 pt]{article}
\usepackage{graphicx}
\usepackage{url}
\usepackage{amsmath}
\usepackage{amssymb}
\usepackage{bm}
\renewcommand {\vec}[1] {\ensuremath{\boldsymbol{#1}}}
\usepackage{hyperref}
\hypersetup{pdfstartview={XYZ null null 1.00} }
\hypersetup{pdfpagemode=UseNone}
\title{Stability of 2-body orbits\\ in retarded gravitation theory (RGT)}

\author{C. K. Raju\\AlBukhary International University\\05200 Alor Setar, Malaysia \\\url{ckr@ckraju.net},  \url{ckraju@aiu.edu.my}}

\begin{document}
\maketitle
\begin{abstract}
The recently formulated retarded gravitation theory (RGT) explains the non-Newtonian velocities of stars in spiral galaxies, \textit{without any new hypothesis}, and may hence be tested even in the laboratory. However, doubts have been expressed that those higher rotation velocities in RGT may be due to instabilities. We resolve these doubts by solving the full functional differential equations of RGT for a model 2-body planetary system. The solution is stable and closely agrees with the Newtonian solution for this planetary case. Thus, the big difference between RGT and Newtonian gravity for a spiral galaxy is not due to any instability in RGT.
%04.50.Kd General relativity and gravitation: Modified theories of gravity
%04.80.Cc Experimental tests of gravitational theories  
%02.30.Ks Mathematical methods in physics: delay and functional equations
%98.62.Dm Kinematics, dynamics, and rotation of galaxies
\end{abstract}
%\keywords {Retarded gravity, galactic rotation curves, flyby anomaly, functional differential equations, experimental tests of gravity}
%\classification{04.50.Kd, 98.62.Dm, 04.80.Cc,02.30.Ks}

\section{Introduction}

\subsection{Preliminaries}

Recent evidence (from supernovae\cite{Riess, Perlmutter}, and the cosmic microwave background 
\cite{Dunkley, Komatsu, Sherwin, Engelen}) 
indicates that the expansion of the cosmos is accelerating. This has belied long-standing expectations, based on Newtonian gravity or general relativity theory (GRT), that the expansion must decelerate. This acceleration can, of course, be explained by populating the cosmos with a mysterious invisible substance which exerts negative pressure---dark energy. 
However, this has also led to doubts about the validity of our current understanding of gravitation.

Another departure from Newtonian expectations has long been known: the rotation velocities of stars in spiral galaxies \textit{increase} as we move out to the edge of the galaxy, instead of decreasing, as expected on Newtonian gravity.\cite{RubinA} This departure too has been explained by populating the galaxy with invisible dark matter. A further fact contrary to Newtonian expectations is that the observed rotation velocities of stars reach a \textit{constant} value at the edge of the galaxy.\cite{Rubin-Sofue} But this too can be explained by advancing yet another hypothesis that the dark matter is distributed in the form of a halo, with its  density peaking where the luminous matter thins out to nothing. Furthermore, this is not just some unseen mass, but specifically cold dark matter---involving some hypothetical form of particles for which there is no laboratory evidence. This has led to the skepticism that the ``standard model'' of cosmology ``$\Lambda$CDM\dots relies heavily on two potentially fictitious invisible entities''.\cite{Famaey}
A theory ceases to be scientific (refutable) if it counters every observational surprise with a new hypothesis, and its criticism is inhibited on grounds of ``standardised''  community opinion. The other common alternative is modified Newtonian dynamics (MOND)\cite{MilgromB, Famaey} 
which modifies the gravitational force at scales beyond a ``characteristic scale'', but in a purely phenomenological way, and goes on to accumulate further hypotheses.

This suggests the need for a theoretically rigorous approach, where such speculative hypotheses (whether about dark matter or about scale-driven changes in the gravitational force) are avoided. 
Accordingly, a new explanation has recently been advanced,\cite{ckr-RGT} for galactic rotation curves, \textit{without any new hypothesis}. The non-Newtonian velocities in spiral galaxies may be just a consequence of Lorentz covariance (which itself is a theoretically  \textit{essential} correction\cite{ckr-Titcon} to Newtonian physics).

This seems astonishing at first sight, for special relativistic ($\frac{v}{c}$) effects are believed to be relevant only at relativistic velocities $v \approx c$. However, while that is definitely true for the \textit{one} body problem, does it remain true for a billion-body problem? Even when $v \ll c$ could a systematic $\frac{v}{c}$ effect become significant when summed over a billion bodies or more, such as the stars in a spiral galaxy? 

This question cannot be answered using GRT (or similar geometrical theories), since with GRT one is  barely able to do the 2-body problem after a century. Nor can it be answered with the parametrized post-Newtonian (PPN) formulation (whether regarded as an approximation to GRT or as an independent theory). For a billion-body problem, Newtonian gravitation was, until recently,  the only available option, and that obviously cannot be used to answer the question whether $\frac{v}{c}$ effects can accumulate when the number of bodies is large. However, the question can now be answered (and the answer is yes) within retarded gravitation theory (RGT).\cite{ckr-RGT}

\subsection{Retarded gravitation theory}

RGT is a Lorentz covariant theory of gravitation which agrees with Newtonian gravitation in the \textit{static} case.  It was recently formulated as follows. 
Consider two particles, with world lines given by   $Y_1 (s_1)$ and $Y_2 (s_2)$, where $Y_1$ and $Y_2$ are 4-vectors, and $s_1$ and $s_2$ are the respective proper times. 
The equations of motion in RGT are
\begin{equation}
	m_1 \frac{d^2Y_1}{ds_1^2} = F_{12} , \quad
	m_2 \frac{d^2Y_2}{ds_2^2} = F_{21} ,
\label {fulleq}
\end{equation}
where $m_1$ and $m_2$ are the respective rest masses of the two particles, and $F_{12}$ the 4-force exerted by particle~2 on particle~1 is given by the Lorentz covariant expression
\begin{align}
	F_{12} &= -\frac{kc^3}{(R_{2\, \text{ret}}.V_{2\, \text{ret}})^3} R_{2\, \text{ret}} + \frac{kc^3}{(R_{2\, \text{ret}}.V_{2\, \text{ret}})^3}\frac{(R_{2\, \text{ret}}.V_1)}{(V_{2\, \text{ret}}.V_1)} V_{2\, \text{ret}} \label{numbered1} \\
	&\equiv \left [ -\frac{kc^3}{(R_2.V_2)^3} R_2 + \frac{kc^3}{(R_2.V_2)^3}\frac{(R_2.V_1)}{(V_2.V_1)} V_2  \right ]_{2\, \text{ret}} .
	\label{numbered2}
\end{align}
Here, $k = Gm_1 m_2$, $G$ is the Newtonian gravitational constant, $c$ is the speed of light, $R_{2 \, \text{ret}} = Y_{2 \, \text{ret}} - Y_1$ is the retardation vector,   $V_1 = \frac{dY_1}{ds_1}$ and $V_2=\frac{dY_2}{ds_2}$ denote the respective 4-velocities, and, in \eqref{numbered2}, $\left [ \, \right ]_{2 \, \text{ret}}$ indicates that the quantities with subscript 2 are to be evaluated at the corresponding retarded proper time, as explicitly indicated in \eqref{numbered1}.  The other force $F_{21}$ is given by interchanging 1 and 2 in \eqref{numbered2}. 

In coordinates, if $Y_1 = (ct, \vec{y}_1(t))$, and $Y_2 = (ct, \vec{y}_2(t))$, the retarded coordinate time $t_{12}$, in the force $F_{12}$ acting on $Y_1$ at time $t_0$, is the root of the equation 
\begin{equation}
	c^2 (t-t_0)^2 = r_{12}^2 \equiv (\vec{y}_2 (t)-\vec{y}_1 (t_0))^2,
	\label{ret-time} 
\end{equation}
satisfying $t < t_0$. That is, it is the value of $t$ at the spacetime point where the backward null cone from $Y_1(t_0)$ intersects the world line $Y_2$. The corresponding distance $r_{12}$ is the retarded distance from particle~1 to particle~2. A similar equation holds for $t_{21}$, the retarded coordinate time in $F_{21}$, the asymmetry being only in the arguments of $\vec{y}_1$ and $\vec{y}_2$.

For non-relativistic velocities, the force in RGT approximately reduces to the simpler expression 
\begin{equation}
F_{12} \approx \frac{k}{r_{12}^2} \left (\frac{R_{2 \, \text{ret}}}{r_{12}} + \frac{V_{2 \, \text{ret}}}{c} 
\right ) ,
\label{approx}
\end{equation}
which still differs from the Newtonian gravitational force by its (tiny) velocity dependence. (It is exactly the Newtonian gravitational force, if the two masses are relatively at rest.)

In a spiral galaxy, all stars in the neighbourhood of a given star are rotating in the same direction.
Though the rotational velocities are  non-relativistic ($\frac{v}{c} \sim 10^{-4} \ll 1$), the cumulative effect of the $v/c$ term, summed over $10^{10}$ co-rotating stars, becomes substantial. Hence, on RGT,  increasing rotation velocities are \textit{expected} in a spiral galaxy. While dark matter might exist, there is no need in RGT to hypothesize dark matter just to explain non-Newtonian rotation velocities in spiral galaxies. Nor is it necessary to make the further hypothesis that the dark matter is peculiarly distributed in the form of a halo, just to explain the flattening of the rotation curves. It is to be expected on RGT that, at the edge of the galaxy, where the effects of the central mass diminish, the  velocity-dependent forces between neighbouring stars,   will make the relative velocities go to zero (i.e., the rotation curves will flatten out).  Since RGT is based on Lorentz covariance alone,  this explains the non-Newtonian behaviour of galactic rotation curves in a way which is \textit{simpler} (i.e., makes no hypothesis), and hence to be preferred over both dark matter and MOND.

Since RGT makes no hypothesis, the refutable consequences of RGT are not confined to astrophysics: RGT can be tested in the laboratory. In a modified Cavendish experiment, if the larger (attracting) masses are rotating this would result in a torque slightly different from the one when they are static. This also means that static and dynamic ways of measuring the Newtonian gravitational constant\cite{NewtonG} $G$ may lead to different values.  
The effects of RGT could also be tested by carefully-designed near-earth flyby of spacecraft which would be affected by the rotation of the earth  as a $\frac{v}{c}$ effect\cite{Anderson2008} along the lines calculated earlier.

We emphasize that RGT (which requires only Lorentz covariance) remains useful regardless of how it compares with GRT (which demands general covariance).  Thus, GRT is simply not usable in practice in situations (like the galaxy) where what is used instead is  Newtonian gravitation, and RGT (since it is Lorentz covariant) is a definite improvement on \textit{that}. Therefore, RGT remains the theory of choice for numerous-body problems. Further, RGT, rather than Newtonian gravitation, remains the correct flat spacetime limit to use.

\subsection{Arguments against retardation}

Nevertheless, there have been objections to retardation in gravity, since pre-relativistic times, when naive theories of retarded gravity were first proposed (to explain the discrepancy in the perihelion advance of Mercury).
A key objection to those naive theories rests on an argument mentioned by Eddington (which we will call ``Eddington's'' argument). The argument is  that two-body orbits would not be stable with retarded forces.\cite{Eddington}
 
The argument assumes that the two bodies were initially rotating rigidly about a common centre of mass. It further assumes that the retarded gravitational force must point to the retarded position of the ``attracting'' body, so that its line of action would not pass through the instantaneous centre of mass. On Newtonian mechanics, this means that angular momentum would not be conserved. Hence the conclusion that 2-body orbits would be unstable unless the ``speed of propagation of gravity'' is very large, as first asserted by Laplace.\cite{Laplace} 
This argument \textit{does} apply to naive theories of retarded gravity, as has been argued in the past,\cite{Good} and as we show more rigorously in this article. 

However, similar arguments have also been raised against both GRT\cite{van-Flandern} and RGT (in personal communications), and this article aims to show that ``Eddington's'' argument does \textit{not} apply to RGT. 
Now, it is problematic to define the centre of mass in GRT (or even in special relativity\cite{Pryce}). Further, there is no physical way for one particle to know the instantaneous position of the other, in \textit{any} relativistic theory.
Does that mean that requiring even Lorentz covariance makes all 2-body orbits unstable?  
In defence of GRT and PPN it has been argued,\cite{Carlip}
using Kinnerseley's ``photon rocket'', that the  velocity-dependent terms in the $\Gamma^\mu_{\nu \sigma}$ cancel the deviation due to retardation, so that the net acceleration points to the ``instantaneous'' centre of mass up to terms of order $\frac{v^3}{c^3}$, and the effect of the inexact cancellation shows up only in the slowly decaying orbits of binary pulsars. 

\subsection{FDEs of RGT}

One can use a similar approach in RGT. 
However, to proceed in a theoretically more rigorous way, we first need to take into account a key new feature of RGT. 
The RGT equations of motion \eqref{fulleq} are retarded \textit{functional differential equations} (FDEs), or state-dependent delay differential equations. Hence, the criterion of stability used in ``Eddington's argument'' does \textit{not} extend automatically to RGT, since retarded FDEs are fundamentally different from ODEs.\cite{ckr-Titcon}
Even when the retardations are tiny, and seem physically negligible, \textit{it is incorrect to approximate retarded FDEs by ODEs}, using a ``Taylor'' expansion in powers of the delay, and this can result in artificial instability, no matter how tiny the delay, so long as it is non-zero.\footnote{For a worked-out counter-example, see \cite{ckr-fde-1}.} 

Thus, to examine the problem of stability in RGT in a theoretically rigorous way, we first need to solve the FDEs of RGT. 

\subsection{Aim} 

Accordingly, here we report on a solution of the FDEs of the 2-body problem in RGT. (The solutions reported earlier for both the galaxy and NASA spacecraft were essentially 1-body solutions for a body in the presence of a relatively large rotating mass, since the aim there was only to bring out the effects of the velocity dependent term in the RGT force, in those situations, where the minute movement of the larger body (earth, galaxy) due to the smaller one (spacecraft, star) is of no concern and can be neglected.) 
Our aim, in this article, is limited to the question of \textit{stability} of 2-body orbits in RGT, when the full FDEs are solved. This also clarifies exactly how to solve the full FDEs of RGT. Other issues will be examined in subsequent articles. 

\section{FDEs of motion in RGT}

Since the two equations \eqref{fulleq} have to be solved simultaneously, it is convenient to use a common time parameter, which we take to be the coordinate time $t$. We assume that the functions $t=t_1(s_1)$ and $t=t_2(s_2)$ are suitably invertible and (at least) twice continuously differentiable, and will not explicitly indicate them further. Thus, we have 
$\frac{dt}{ds_1} = \gamma_1$, and $\frac{dt}{ds_2} = \gamma_2$,
where $\gamma_1$ and $\gamma_2$ are the respective Lorentz factors. Using an overdot to denote derivatives with respect to $t$, we have, by the chain rule,  
$V_1 = \frac{d Y_1}{ds_1} = \frac{dY_1}{dt}\frac{dt}{ds_1} = \gamma_1 \dot{Y_1}.$
Similarly, 
$\frac{d V_1}{ds_1} = \frac{dV_1}{dt}\frac{dt}{ds_1} = \gamma_1 \dot{V_1} .$

Hence, 
\eqref{fulleq} can be rewritten
\begin{align}
	\dot{Y}_1 &= \frac{1}{\gamma_1} V_1 , \nonumber\\
	\dot{V}_1 &= \frac{1}{\gamma_1} \frac{F_{12}}{m_1} , 
\end{align}
with similar equations for particle 2.

Since the zeroth component of these equations is not independent, we can write them in 3-vector notation using  $Y_1 = (ct, \vec{y}_1(t))$, $Y_2 = (ct, \vec{y}_2(t))$, 
so that $\dot{Y_1} = (c, \vec{v}_1)$, 
 $\dot{Y_2} = (c, \vec{v}_2)$.  
Let $\vec{u}_1$ and $\vec{u}_2$ denote the space components of the velocity 4-vectors $V_1$, and $V_2$, so that $\vec{u}_1 = \gamma_1 \vec{v}_1$, $\vec{u}_2 = \gamma_2 \vec{v}_2$. Further, we let $\vec{r}_{2 \, \text{ret}} = \vec{y}_2({t_{12}}) - \vec{y}_1(t)$, denote the 3-vector corresponding to $R_{2 \, \text{ret}}$. Then the final equations are 
\begin{align}
	\dot{\vec{y}}_1 &= \frac{1}{\gamma_1} \vec{u}_1 \nonumber \\
	\dot{\vec{u}}_1 &= \frac{1}{m_1 \gamma_1} \vec{f}_{12}
	\label{finaleq}
\end{align}
where
\begin{equation}
	\vec{f}_{12} = a \, \vec{r}_{2\, \text{ret}}	+ b \, \vec{u}_{2\, \text{ret}}
	\label{little-f}
\end{equation}
and 
\begin{equation}
a = - \left[ \frac{kc^3}{(R_2 \cdot V_2)^3} \right ]_{2\, \text{ret}} \quad b = a \tilde{b} \quad \tilde{b} =   -\left[ \frac{(R_2 \cdot V_1)}{(V_2 \cdot V_1)} \right ]_{2\, \text{ret}}
\label{ab}
\end{equation}
The equations \eqref{finaleq} (accompanied by \eqref{little-f}, \eqref{ab}), together with the corresponding equations for particle 2 are the four 3-vector equations (or 12 equations in all) we will actually solve. 

\subsection{Past data}

Past data must be prescribed for the solution of the FDEs of RGT (unlike the ``initial'' data appropriate for Newtonian ODEs). We follow our earlier strategy of taking the past data to be given by the theoretical 2-body orbits of Newtonian gravitation, around the (Newtonian) instantaneous centre of mass. Of course, past data could be prescribed in an infinity of other ways, but this strategy suits our immediate purpose, since it allows us to compare RGT with Newtonian gravitation (which, we know, gives stable orbits in the 2-body case).

Since our aim is limited to the question of stability, we take up only the case of the (mean) Sun-Jupiter system (ignoring other planets). The corresponding theoretical, 2-body  \textit{circular} motion is determined in Newtonian gravitation by prescribing the mass ratio of Jupiter (=0.000954265748 solar masses), and its mean distance (=5.2 AU). With our strategy of using the Newtonian motion as the past data, the same parameters also suffice to obtain a solution of  the 2-body problem of RGT.

\subsection{Units}

As explained in earlier publications, efficient computational units must be appropriate to the past data, and we will use as units, 1 solar mass, 1 AU, and 1 earth year.

\subsection{Method of solution}

The technique of solving FDEs has been discussed exhaustively in previous publications,\cite{ckr-2004, PaTomu}
and there is nothing new to add.

The planetary 2-body problem in RGT is not stiff, so the code \textsc{retard}\cite{Hairer} is appropriate. When proceeding on the method of steps\cite{Elsgolts1} (used to solve retarded FDEs) the code must be constrained to steps of the size of the delay. The delay ($\approx 8.3 \times 10^{-4}$ years)  
is small compared to the time period ($\approx 100$ years) over which we need to solve the system of equations,  to be able to resolve the question of stability. 
Nevertheless, while codes like \textsc{radar} 
for stiff equations  can take step sizes larger than the delay, they would be needed only in extreme relativistic situations.
For the present, using the simpler and long-established \textsc{retard} code helps to keep the whole calculation transparent.

\section{Results and discussion}

We solved the equations for about 3 centuries, or about 25 Jovian cycles. The motion remains stable and the distance between the two bodies differ by at most  $2 \times 10^{-8}$ AU ($\sim$3 km) from the instantaneous distance between them in the theoretical (circular) Newtonian motion. The marginal difference arises because the orbits do not remain circular but turn into ellipses in RGT (as would happen due to any perturbation on the Newtonian theory). Had we prescribed elliptic rather than circular past orbits, there would no longer be any \textit{visible} difference between the orbits with RGT and Newtonian gravitation.  
Thus, unlike the galactic case, for the planetary case, RGT agree closely with Newtonian gravitation. 

\begin{figure}[htbp]
\includegraphics[width=42pc]{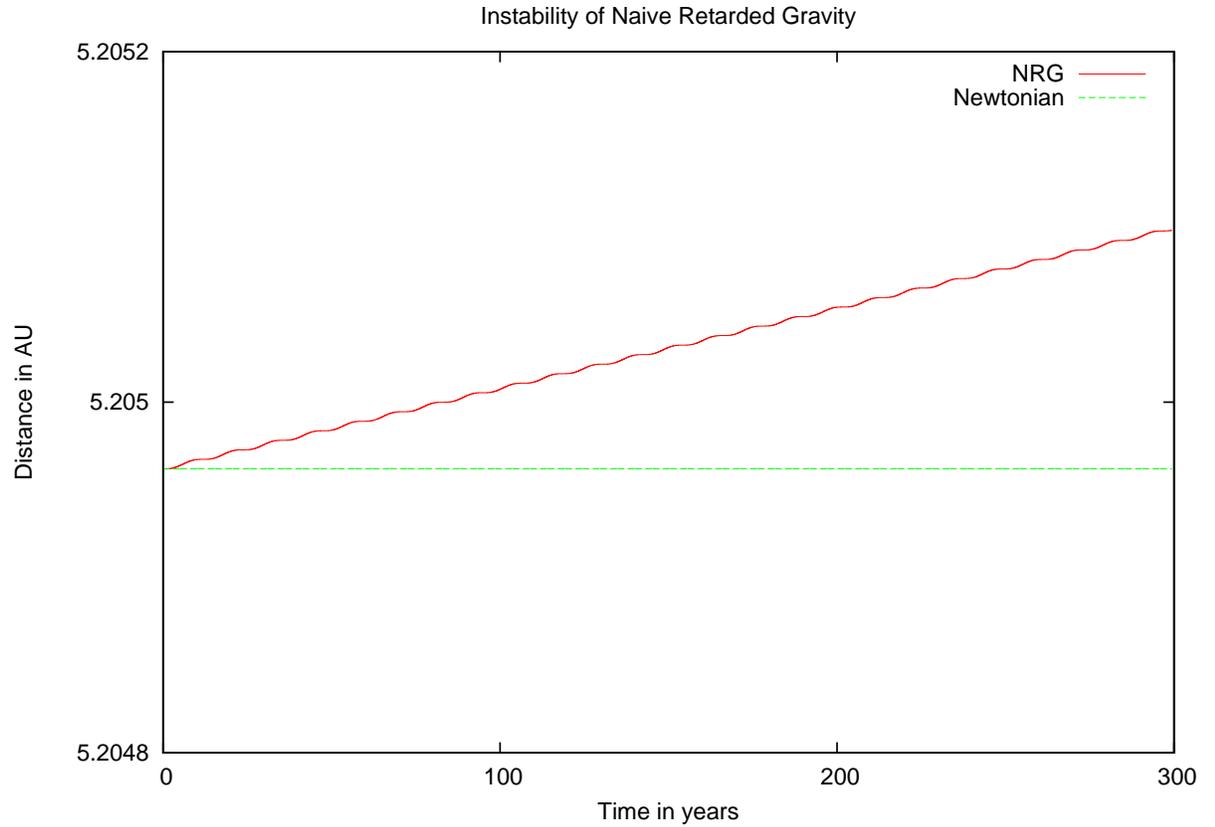}
\caption{In naive retarded gravity (NRG) the force varies inversely as the square of the retarded distance and points in the direction of the retarded position of the other body. The instability of the naive theory is brought out by  the plot which shows how the relative distance between the two bodies ($\sqrt{(\vec{y}_1 (t) - \vec{y}_2 (t))^2}$) systematically increases over time in the naive theory.}
\label{fig:nrg}
\end{figure}

Specifically, the calculation shows that the delay torque does \textit{not} induce any instability in RGT.
We can try to understand this in Newtonian terms. The explicit velocity dependent ($\frac{v}{c}$) term in the RGT force is essential for its Lorentz covariance. On Newtonian intuition, this term makes the net force point closer to the ``instantaneous centre of mass'' and ensures stability. Suppose that term were absent, as happens in ``naive retarded gravity'' (NRG),  where the force varies inversely as the square of the retarded distance, and points exactly to the retarded position of the other body. In that case of NRG, or a ``retarded inverse square law'',  there is indeed a small instability (Fig.~\ref{fig:nrg}). 

The Lorentz gamma factors are an additional detail to be taken into account. They were absent in ``Eddington's'' argument. These Lorentz factors slow down the velocity of each body by a tiny factor of $\frac{v^2}{c^2}$---but, then, in RGT (unlike NRG) the delay torque too is at most of that order, for the leading order $\frac{v}{c}$ term cancels. As is well-known, the perihelion shift of Mercury is an effect of order $\frac{v^2}{c^2}$, but those tiny differences for the case of elliptic orbits require a detailed analysis, and we report on them separately.  

Establishing the stability of planetary motion in RGT sets to rest the doubt that the higher-than-Newtonian velocities obtained using RGT for the case of galaxy were due to instability. Due to the different configuration (of masses), and prescribed past data in the two cases, the explicit velocity dependent ($\frac{v}{c}$) term, in the RGT force, systematically adds up in the galactic case, while, in the planetary case, it cancels to ensure stability. Hence, RGT agrees with Newtonian gravity for planetary motion despite giving significantly different results for the galaxy. 

Establishing that the different results for the galaxy are not due to any intrinsic instability in RGT opens the way to explore the possibility that the accelerating expansion of the cosmos may be similarly due to a different configuration (such as cosmic rotation) rather than yet another invisible new substance.

\section{Conclusions}

We have solved the full FDEs of RGT for a model 2-body planetary system. The system remains stable, and RGT gives results almost identical to Newtonian gravity for this case of planetary orbits. Thus, the different results that RGT gives for the galaxy are not due to any instability. 

\section*{Acknowledgment} The author is grateful to Dr S. Raju for discussions, and to Prof. S. Carlip for comments on an earlier draft.

%\pagebreak

\begin{figure}[htbp]
	\includegraphics[width=42pc]{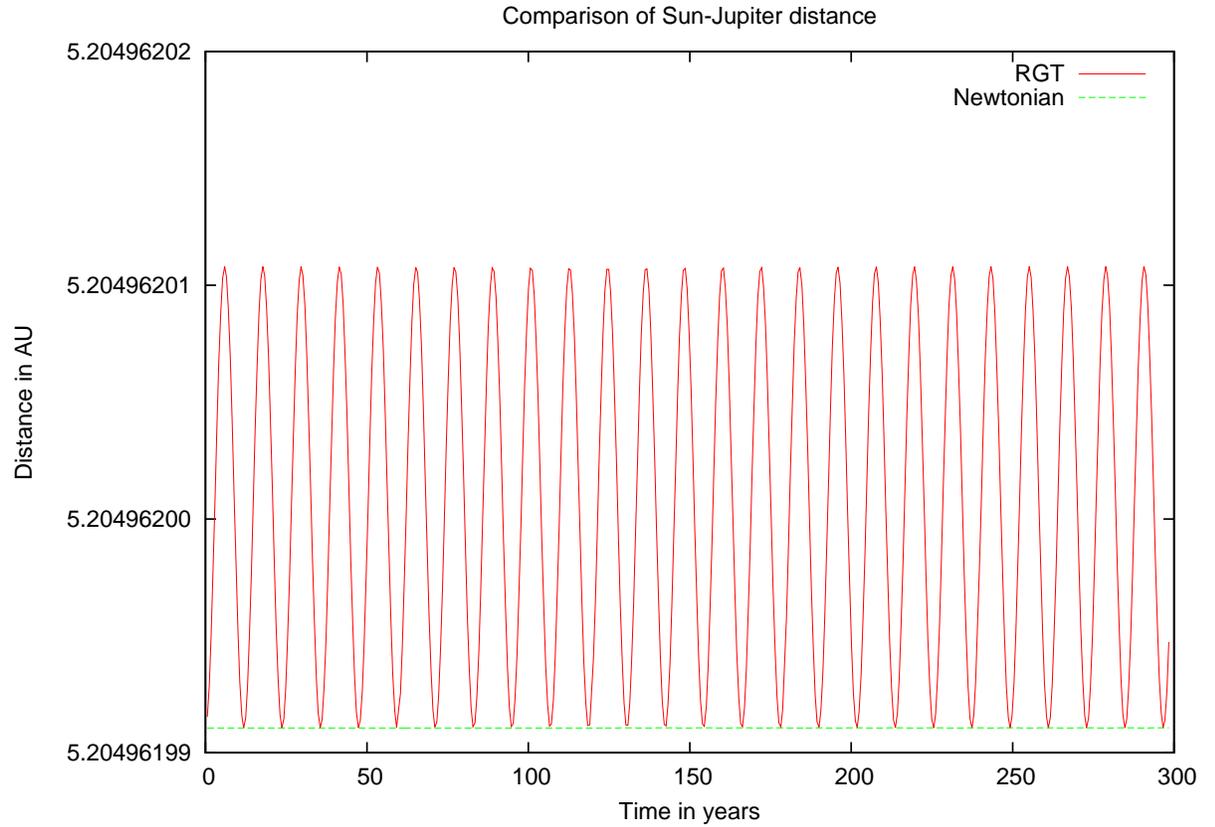}
	\caption{The relative distance of the two bodies ($\sqrt{(\vec{y}_1 (t) - \vec{y}_2 (t))^2}$) stays stable. The RGT distance differs by at most $2\times 10^{-8}$ AU or $\sim 3$ km from the Newtonian distance. This difference arises since RGT changes the prescribed Newtonian circular orbit to an ellipse.}
	\label{fig:RelativeDistance}
\end{figure}

\begin{figure}[htbp]
	\includegraphics[width=42pc]{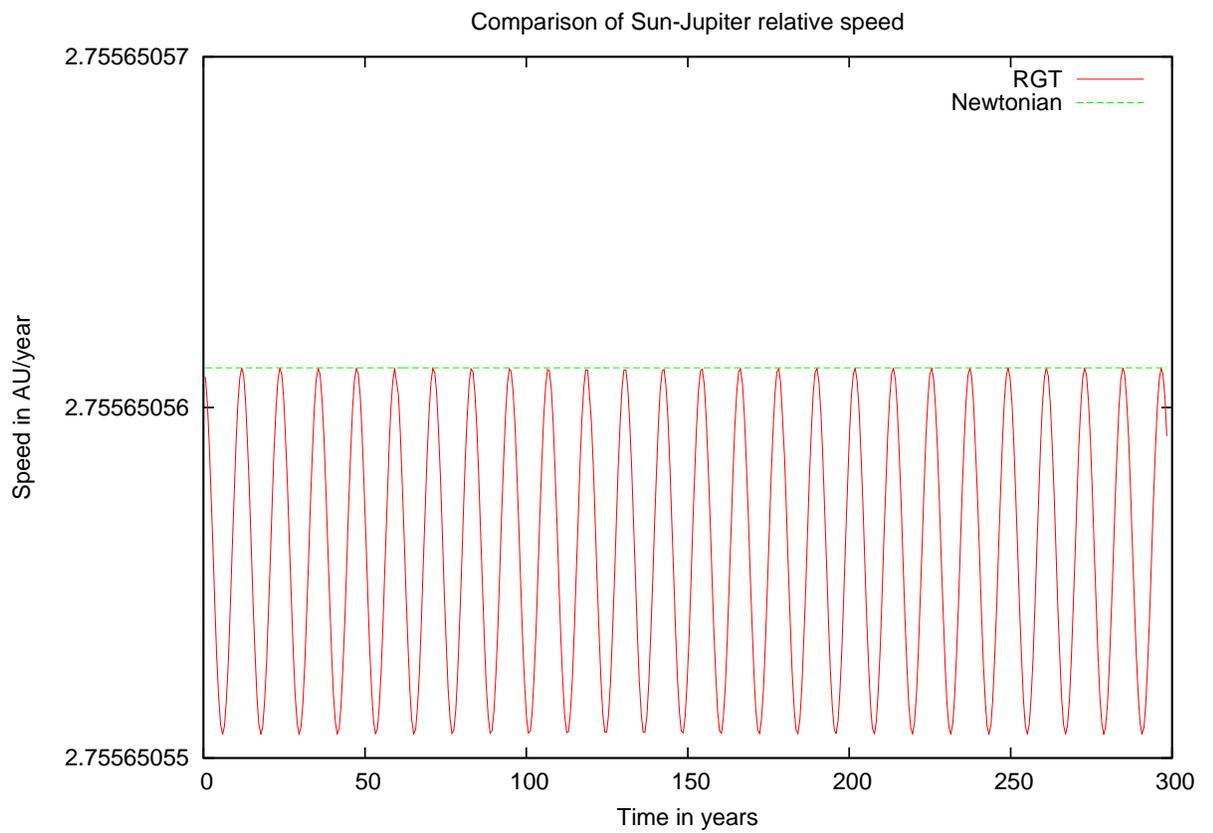}
	\caption{The relative speed of the two bodies ($\sqrt {(\vec{v}_1 - \vec{v}_2)^2}$) remains stable. The maximum difference from the Newtonian case is $10^{-8}$ AU/year or around $4.74 \times 10^{-8}$ km/s.}
	\label{fig:RelativeSpeed}
\end{figure}

\pagebreak
\bibliographystyle{unsrt}
%\bibliography{refFDE}

\end{document}